# Impacts of Doping on Epitaxial Germanium Thin Film Quality and Si-Ge Interdiffusion

Guangnan Zhou[1], Kwang Hong Lee[2], Dalaver H. Anjum[3], Qiang Zhang[4], Xixiang Zhang[4], Chuan Seng Tan[2] and Guangrui (Maggie) Xia[1,*]

[1]Department of Materials Engineering, University of British Columbia, 309-6350 Stores Rd, Vancouver, BC V6T1Z4, Canada.
[2]Low Energy Electronic Systems (LEES), Singapore-MIT Alliance for Research and Technology (SMART), 1 CREATE Way, #10-01 CREATE Tower Singapore 138602.
[3]Imaging and Characterization Core Lab, King Abdullah University of Science and Technology, Thuwal 23955-6900, Saudi Arabia
[4]Division of Physical Science and Engineering, King Abdullah University of Science and Technology, Thuwal 23955-6900, Saudi Arabia

**Abstract**

Ge-on-Si structures with three different dopants (P, As and B) and those without intentional doping were grown and annealed. Several different materials characterization methods have been performed to characterize the Ge film quality. All samples have a smooth surface (roughness < 1.5 nm), and the Ge films are almost entirely relaxed. On the other hand, B doped Ge films have threading dislocations above $1 \times 10^8$ cm$^{-2}$. While P and As doping can reduce the threading dislocation density to be less than $10^6$ cm$^{-2}$ without annealing. The interdiffusion of Si and Ge of different films have been investigated experimentally and theoretically. A quantitative model of Si-Ge interdiffusion under extrinsic conditions across the full $x_{Ge}$ range and with the dislocation mediated diffusion term was established. The Kirkendall effect has been observed. The results are of technical significance for the structure, doping, and process design of Ge-on-Si based devices, especially for photonic applications.

## 1. Introduction

An unprecedented technology boom has occurred in silicon photonics in the recent two decades. Germanium (Ge), as the most silicon (Si) compatible semiconductor, are playing an increasingly important role in large-scale dense Si photonic integration, such as in light sensing and modulation [1, 2]. In the past few decades, researchers all over the world have invested extensive efforts in finding solutions to a Si-compatible lasing material system [3-9]. Among all of the candidates that have been extensively researched, III-V quantum dot (QD) lasers grown on Ge-on-Si substrates and Ge-on-Si lasers [10-12] have been demonstrated to be among the most promising on-chip light sources [13]. Besides laser applications, there are other Ge-on-Si structure-based devices, such as SiGe modulators, Ge photodiodes and GaAs-based devices on Ge/Si substrates, where different types of dopants can be involved.

It is important to consider the Ge film functions and quality requirements for different optoelectronic devices with Ge-on-Si structures. The greatest challenge for high-quality Ge epitaxy on Si is the 4.2% lattice mismatch between the two materials. This mismatch causes two serious issues: high surface roughness resulting from the Stranski–Krastanov growth, and a high density of threading dislocations (TDs) in Ge epitaxial layers. High surface roughness and high density of TDs both severely affect the performance of Ge photodiodes and lasers because of the recombination centers that are introduced along these dislocations [1]. For Ge MOSFETs, dislocations reduce carrier mobility and cause high leakage current.

It is also very critical to minimize Si-Ge interdiffusion in Ge-on-Si lasers, optical modulators and other Ge-on-Si based optoelectronic devices as interdiffusion changes Ge profiles and all properties related to Ge concentration such as the bandgap, effective mass, carrier mobility, carrier lifetime etc. According to the Ge laser simulations work of Li *et al.* [14] and Ke *et al.* [15], it is clear that Si-Ge interdiffusion is one of the key reasons for the low efficiency and high threshold current density. From the prototype structure of Ge laser [12], the thickness of the Ge layer is on the order of $10^2$ nm, which is thin enough to be susceptible to Si-Ge interdiffusion. Therefore, it is important to understand Si-Ge interdiffusion in Ge-on-Si structures.

Despite its increasingly important role in photonic integration, there is a substantial lack of studies on the doping impact on Ge-on-Si film quality. Lee *et al.* studied the impact of high concentration arsenic (As) on Ge epitaxial film grown on Si (001) with 6° off-cut. He concluded that the TDs density had been reduced by at least one order of magnitude to $< 5 \times 10^6$/cm$^2$ and attributed that to the enhancement in the velocity of the dislocation motion in an As-doped Ge film [16]. This is a promising method to produce n-type doped Ge for Ge-on-Si lasers or high quality

Ge as a transition layer for GaAs and Si integration.

There have been a handful of studies on Si-Ge interdiffusion with/without doping. Xia et al. first used Boltzmann-Matano method to study Si-Ge interdiffusion and studied a few impacting factors of interdiffusion including temperature, tensile strain, compressive strain, Ge concentration, and oxidation [17-19]. Dong *et al.* [20] established a benchmarking model for Si-Ge interdiffusivity over the full Ge fraction range based on Darken's law and thermodynamics theory, which agrees with the vast majority of the experimental data in the area. In 2002, Takeuchi and Ranade *et al.* [21, 22] studied the Ge-Si interdiffusion in a polycrystalline Ge/Si structure under As doping. They reported that the interdiffusion was enhanced by about five times when the As doping level was $1 \times 10^{21}$ cm$^{-3}$. In 2008, Gavelle *et al.* [23] studied the impact of boron (B) on Si-Ge interdiffusion in a Ge on Si structure. The interdiffusion is retarded when the Ge layer is doped with boron. On the other hand, Ranade *et al.* [21] reported that Si-Ge interdiffusion had been enhanced with boron doping in 2002. Our group recently showed that high phosphorus (P) doping greatly accelerates Si-Ge interdiffusion due to the Fermi-level effect [24, 25]. Cai *et al.* [25] successfully established a quantitative model of interdiffusion with that theory. However, only one dopant (P) was involved and the Ge fraction was limited to 0.75 < $x_{Ge}$ < 1. The accuracy of this model requires more experimental data to fine-tune and more data were needed for the full Ge range. Especially, interdiffusion with other dopants had not been well studied systematically, which was addressed in this work.

Ge-on-Si layers form Ge/Si interdiffusion couples. Another task that motivated this study was to observe the Kirkendall effect, which exists for interdiffusion couples formed between two different materials with different intrinsic diffusivities and is the basis for Boltzmann-Matano analysis and thus the interdiffusivity extraction from interdiffused profiles. Historically, to observe the Kirkendall effect, inner marker layers were used. However, the interdiffusion for the device applications is in nanometer scale and in high quality Ge/Si systems. The addition of marker layers should not degrade the Ge quality requirement or device performance, which is hard to achieve. The Kirkendall effect is about the move of the lattice sites from one side of the interdiffusion couple to the other side. Due to thickness and depth uncertainly during epitaxial growth and profiling, although interdiffusion can be easily measured, the Kirkendall effect had not been observed previously before this work.

In this work, Ge-on-Si structures with three different dopants (P, As, B) and without dopants were grown and annealed. Several different materials characterization methods have been performed to characterize the film quality. The interdiffusion of Si and Ge of different films have been investigated experimentally and theoretically across the whole Ge molar fraction ($x_{Ge}$) range. It is also the first time that Kirkendall effect has been observed in Si-Ge interdiffusion system.

## 2. Impacts of doping on Ge film quality
### 2.1 Structure design, growth and defect annealing

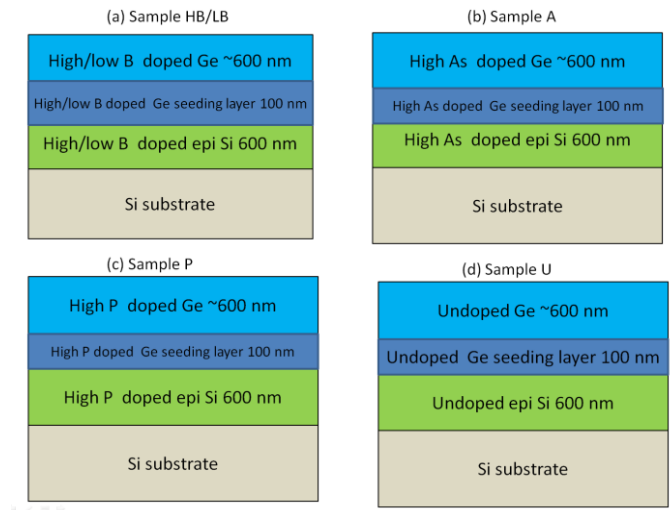

Figure 1 Schematic diagrams of the structures in this work (a) B doped sample with high/low concentration; (b) As doped sample; (c) P doped sample; and (d) undoped sample.

12 Ge-on-Si samples with 5 doping configurations (Figure 1) and 3 annealing conditions (no annealing, 5 thermal cycles, and merged annealing) were designed. Sample U, A, P, B stand for undoped Ge/Si, As doped Ge/Si, P doped Ge/Si, and B doped Ge/Si respectively. Furthermore, we designed two different boron concentrations to study boron doping level and the Fermi level effect for B doped Ge/Si. One is with higher boron concentration (HB), and the other one is with a lower concentration (LB). The B concentration in Sample LB is lower than the intrinsic carrier density of Ge ($n_{i,Ge}$) except for the Ge/Si interface as seen in Figure 8 in Section 3.1. Dopants concentration in Sample P, A and HB are higher than $n_{i,Ge}$. P and As doping levels were chosen as the highest concentration achievable in the epitaxial growth tool.

All samples were grown on 8-inches (100) Czochrolski

(CZ) Si wafers in a metal-organic chemical vapor deposition (MOCVD) tool model CRIUS CCS from Aixtron. The undoped, P and As doped wafers are 6° off-cut towards the [110] direction. As B-doped Ge is too rough for the subsequent materials growth, so we used the on-axis Si wafers for the Ge growth except for the sample LB-5TC as a control sample. The wafer orientation information is listed in Table 1. Before a Si layer was deposited, the Si substrate was treated at 1050 ± 10 °C for 10 minutes under $H_2$ ambient at 400 mbar. Then, a 600 nm doped or undoped Si layer was deposited at 950 ± 10 °C under $H_2$ ambient at 100 mbar. To improve the Ge film quality and reduce the threading dislocations caused by the Ge-Si lattice difference, a 100 nm doped/undoped Ge seeding layer was deposited at 400 ± 10 °C under $H_2$ ambient at 100 mbar (low-temperature Ge growth) on top of the Si layer. Finally, a doped or undoped Ge film about 600 nm was deposited at 650 ± 10 °C under $H_2$ ambient at 100 mbar (high-temperature Ge growth). We denote these layers as "the top Ge layers" in the discussion below to differentiate from the Ge seeding layers. All the dopants quoted were in-situ doped during the growth of the corresponding layers.

Immediately after the growth procedure, half of the samples were annealed inside the growth tool while another half were left unannealed (NA) for comparison. Post-deposition thermal cycling was performed by repeating a hydrogen annealing cycle between low annealing temperature (LT) and high annealing temperature (HT) ranging from 600 °C to 850 °C for 5 times (5×). Each annealing step at HT was 10 minutes, at LT was 5 minutes, and the annealing was performed in an $H_2$ environment to improve the quality of the Ge epitaxial film. The ramping up rate was around 1°C/s and the cooling down rate was also around 1 °C/s. Besides, for Sample LB and Sample P, we performed the merged high temperature (HT), namely, with 850°C for 50 minutes anneal, an isothermal anneal at this temperature with no LT steps or cycling to check the difference between HT/LT thermal cycling and merged HT annealing. The temperatures quoted above were the nominal setting temperatures of the MOCVD reactor. After we obtained the Secondary Ion Mass Spectrometry (SIMS) data and compared with well-established interdiffusion model by Dong *et al.* [20], which was based on many studies in this field, we found that 890 °C is the best fitting temperature using Dong et al.'s model for Sample U/P/LB/HB, and we consider that was a calibrated experimental annealing temperature. For the sample A, since the surface temperature measured was about 20 °C lower than other samples, we believe the calibrated annealing temperature is 870 °C.

**2.2 Roughness characterization**

Atomic Force Microscope (AFM) measurements were performed to obtain the surface roughness information. The scanning size was 1 μm × 1 μm. Several different scanning areas in the samples have been chosen to calculate the surface roughness. The calculation results are listed in Table 1. The roughness of the samples are around 0.3 - 1.5 nm, which depends on the area selected and calculation methods. Different dopant configurations or annealing procedures have no significant effect on the surface roughness. That smooth surface is suitable for the transition layer application between GaAs and Si.

|  | Wafer offcut | Average roughness (nm) | RMS roughness (nm) |
|---|---|---|---|
| U-NA | 6° | 0.65 ± 0.15 | 0.91 ± 0.26 |
| U-5TC | 6° | 0.59 ± 0.20 | 0.80 ± 0.29 |
| P-NA | 6° | 0.42 ± 0.08 | 0.57 ± 0.14 |
| P-5TC | 6° | 1.05 ± 0.35 | 1.49 ± 0.48 |
| A-NA | 6° | 0.34 ± 0.05 | 0.32 ± 0.03 |
| A-5TC | 6° | 0.28 ± 0.01 | 0.35 ± 0.01 |
| A-HT | 6° | 0.82 ± 0.29 | 1.43 ± 0.69 |
| LB-NA | 0° | 0.27 ± 0.04 | 0.34 ± 0.05 |
| LB-5TC | 6° | 0.28 ± 0.03 | 0.36 ± 0.03 |
| LB-HT | 0° | 0.31 ± 0.03 | 0.39 ± 0.04 |
| HB-NA | 0° | 0.52 ± 0.17 | 0.78 ± 0.35 |
| HB-5TC | 0° | 0.66 ± 0.25 | 0.97 ± 0.45 |

Table 1 Wafer offcut information and the average and RMS surface roughness of the samples. The offcut is towards [110].

**2.3 X-ray diffraction**

High resolution X-ray diffraction (HRXRD) measurements were performed to measure the Ge strain level of the samples. All the measurements were performed using a PANalytical X'Pert PRO MRD with a triple axis configuration. Strain values of the Ge layers are extracted by fitting with the PANalytical Epitaxy software package.

The (0 0 4) Ω–2θ scans results of the samples are shown in Figure 2. The position of the Ge peaks of all sam-ples are under biaxial tensile strain in comparison with fully relaxed Ge peaks. For the unannealed samples, the associated degrees of relaxation R calculated by combining Eqs. (3) and (4) from Ref. [26] are within the range 103.9% - 105%. This means that the Ge layers are in a slightly tensile strained configuration (~ 0.16%), which is in agreement with the results reported by Hartmann *et al.* [26, 27] and MIT researchers [28]. The tensile strain is thermally induced to the Ge epilayer during cooling from high-temperature growth or thermal annealing steps to room temperature. In the temperature range of 20 °C to 650 °C, Ge has a coefficient of thermal expansion (CTE) of 5.8 – 8.1 ppm/°C larger than that Si, which is 2.6 – 4.1 ppm/°C [29].

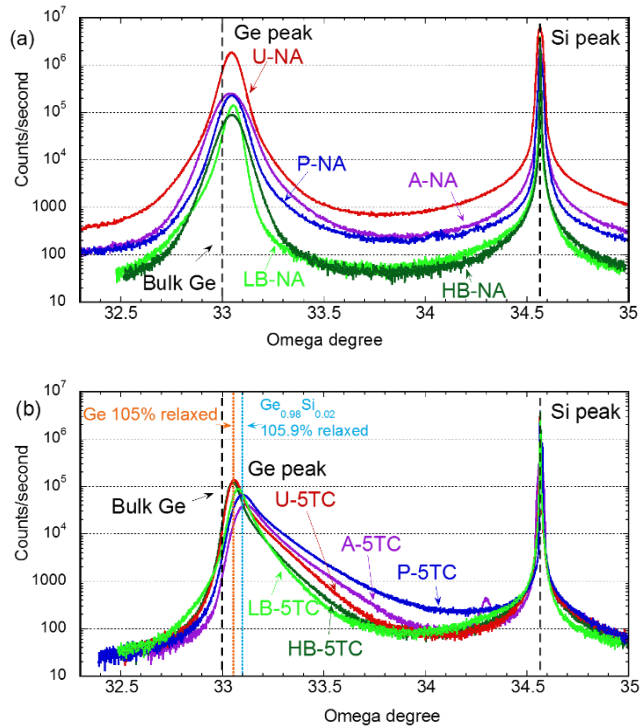

**Figure 2 HRXRD results of the samples (a) without annealing; and (b) after annealing. The results show that the Ge layers are almost fully strained relaxed.**

For the samples after annealing, the HRXRD results are more complicated. Compared with the HRXRD results of the samples before annealing, the Ge peaks of the samples after annealing are much more asymmetric. They are wider towards the high incidence angle side. This is due to SiGe interdiffusion during the annealing, forming a SiGe alloy region with a graded Ge concentration. This is consistent with SIMS results in Section 3. Sample P-5TC and A-5TC have larger interdiffusion than other samples, making their Ge peaks more broadened. The interdiffusion also shifts the "Ge" peaks towards the Si side, as the "Ge" bulk layers are no longer uniform Ge layers with 100% Ge. For example, according to the SIMS data, at the surface of P-5TC, $x_{Ge}$ = 0.98, and $x_{Ge}$ decreases with the depth. In this case, it will be inaccurate to calculate the exact relaxation R of the Ge layers solely through the (0 0 4) scans. The Ge concentration information should also be included. From the XRD "Ge" peak, the relaxation of this layer can be calculated as 105.9% considering it as $Ge_{0.98}Si_{0.02}$. Considering the fact that $x_{Ge}$ decreases with the depth, the relaxation of all other regions is even less than 105.9%. Thus, we can conclude that for P-5TC, the Ge layer became a SiGe alloy region with the surface Ge concentration being $x_{Ge}$ = 0.98. This SiGe alloy region is still almost entirely relaxed. The same argument can be applied to A-5TC as its $x_{Ge}$ at the top is around 98.5% measured by SIMS. The bottom line is that in the annealed samples, the top layer is a Ge rich layer with similar slight tensile strain, which is not going to influence the interdiffusion significantly [17].

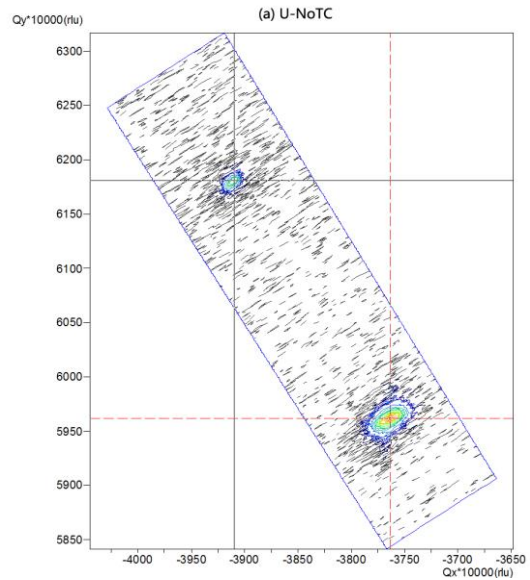

**Figure Error! No text of specified style in document. HRXRD result of (2 2 4) reciprocal space mapping of Sample U-NA.**

To further confirm our results, (2 2 4) reciprocal space maps have been performed for sample U-NA as illustrated in Figure 3. The results are consistent with the conclusion from (0 0 4) Ω–2θ scan. HRXRD results demonstrate that all samples are nearly entirely strain-relaxed. Dopants have no significant influence on the strain degree of the Ge films.

**2.4 Dislocation characterization**

Conventionally, three methods are used to determine the threading dislocation density (TDD) in semiconductor materials: plan-view transmission electron microscopy (PVTEM), cross-section transmission electron microscopy (XTEM) and etch-pit-density (EPD) observation. EPD observation is suitable when TDD is less than $10^7$ cm$^{-2}$, and PVTEM observation is suitable when the TDD is higher than $10^8$ cm$^{-2}$ [30, 31]. Both experiments have been applied to characterize TDD in different samples. Besides, XTEM has been conducted to observe the defects at the Ge/Si interfaces in 2 typical samples.

### 2.4.1 EPD

Firstly, EPD measurements were performed to obtain the TDD of samples. Optical microscope and scanning electron microscope (SEM) imaging were used to observe and count the etch pits. Each sample was etched with iodine ($I_2$) solution. The $I_2$ solution is a mixture of $CH_3COOH$ (100ml), $HNO_3$ (40ml), HF (10ml), $I_2$ (30mg) [32, 33]. The etch rate is approximately 40-80 nm/s depending on the dopants and doping level. After etching roughly half of the top Ge layer (about 300 nm), 4 to 6 different positions on the surface were imaged with an optical microscope and a scanning electron microscope for TDD determination.

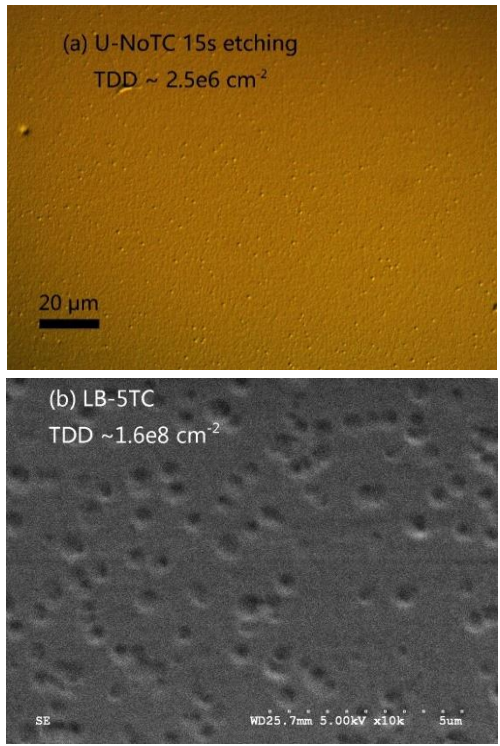

**Figure 4** Example of EPD results of (a) Sample U-NA with 15 s etching; and (b) Sample LB-5TC with 12 s etching. Figure 4(a) was imaged with an optical microscope, and Figure 4(b) with a scanning electron microscope.

Table 2 lists TDD values in the Ge films of the 12 samples measured by EPD. Comparing TDD values of the samples after annealing with different dopants configurations, we can easily found that the boron doped samples have the highest values (> $10^8$ cm$^{-2}$), which means they have the poorest quality. A Ge layer with such dense dislocations is not suitable as a transition layer between GaAs and Si, as the subsequent GaAs growth will be affected. The merged annealing and the 6 degree offcut don't have any impact on the TDD value compared with the corresponding 5TC annealing as seen in LB-5TC (6° offcut) and LB-HT (0° offcut).

For As and P doped samples without annealing, TDDs in sample A-NA and P-NA are in the $10^5$ to $10^6$ cm$^{-2}$ range, one order of magnitude lower than that in U-NA. This TDD level is already low enough for electronic or photonic applications, and there is no need to have an extra defect annealing step for A-NA and P-NA. This also prevents the interdiffusion during the defect annealing step and is one of the major findings of this work. On the other hand, TDDs in A-5TC and P-5TC are higher than that in U-5TC. This can be interpreted by the following pictures:

(1) As previously reported, dislocations move faster in As-doped Ge and slower in Ga-doped than in undoped Ge [34]. This is due to the presence of shallow donor or acceptor levels at the dislocation or other defects such as kinks or antiphase defects. A similar explanation can be applied to P/B. We expect a suppression of dislocation generation in As/P doped Ge and prompting in B doped Ge. Thus, P/A-NA has a lower TDD value, and LB/HB-NA has a higher value than U-NA. (2) In Ge/Si, it is accepted that the dislocation core is a perfect sink for impurity atoms that arrive there [35, 36]. Impurities are known to be effectively gettered by dislocations [37]. Dislocations can be immobilized due to the formation of impurity complexes or clusters at dislocation sites through their accumulation [34]. Thus, P/A-5TC has a higher TDD than U-5TC.

| Sample | TDD value (cm$^{-2}$) | Sample | TDD value (cm$^{-2}$) |
|---|---|---|---|
| U-NA | 3.5 ± 1.5 × 10$^6$ | A-HT | 1.2 ± 0.5 × 10$^5$ |
| U-5TC | < 1 × 10$^5$ | LB-NA | > 2 × 10$^8$ |
| P-NA | 3 ± 1 × 10$^5$ | LB-5TC | 1.2 ± 0.5 × 10$^8$ |
| P-5TC | 1.75 ± 1 × 10$^5$ | LB-HT | 1.1 ± 0.5 × 10$^8$ |
| A-NA | 5 ± 3 × 10$^5$ | HB-No | > 2 × 10$^8$ |
| A-5TC | 1.2 ± 0.5 × 10$^5$ | HB-5TC | > 2 × 10$^8$ |

**Table 2** TDD value of the 12 samples measured by EPD.

In conclusion, boron doping significantly impairs the Ge film quality, while As and P can reduce the TDD level in unannealed samples. This provides a new method to fabricate high-quality Ge-on-Si films without defect annealing procedure, which can avoid undesired Si-Ge interdiffusion.

### 2.4.2 PVTEM

The specimens for PVTEM analysis of samples were prepared by using the chemical wet etching technique in which both HF and HNO3 in 1:1 ratio were utilized perform the back-etching of thinned Si substrates. While the specimens for XTEM were prepared with FIB system of model Helios 450 from Thermo Fisher Scientific. The analysis of both PVTEM as well as XTEM specimens were performed in a microscope of model Titan 80-300 ST also from Therm Fisher Scientific. The microscope was operated at the accelerating voltage of 300 kV during the analysis. PVTEM specimens were tilted to so-called 2beam diffraction before the recording of their images at different magnifications. For the case of XTEM, high-angle annular dark-field (HAADF) scanning TEM (STEM) configuration was utilized to simultaneously enhance contrast from TDs and to suppress diffraction contrast Si lattice. In this way, a clearer images of samples having TDs were realized with XTEM analysis.

Figure 5 shows PVTEM images of the Sample LB-NA and LB-5TC. Some of TDs are indicated by arrows in Figure 5 (a). The TDD values we obtained by PVTEM are larger than the EPD results of the corresponding samples, which is reasonable as PVTEM has a larger magnification.

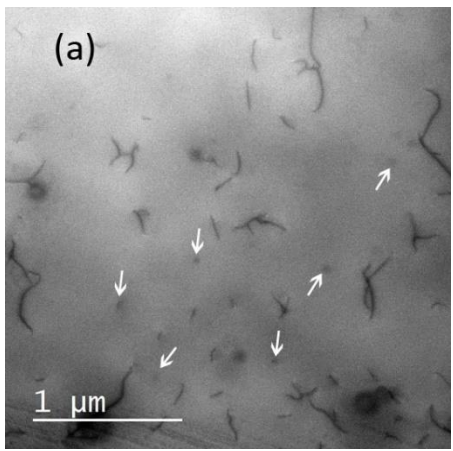

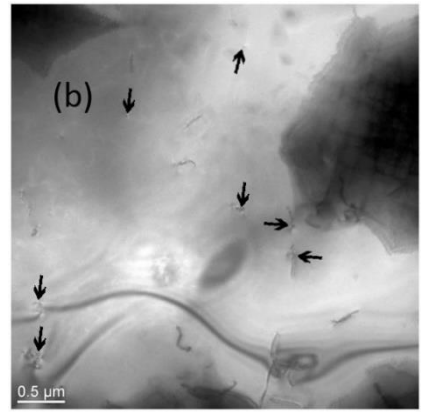

**Figure 5** Images of PVTEM show different shapes and densities of threading dislocations in different samples; (a) LB-NA and; (b) LB-5TC.

### 2.4.3 XTEM

As PVTEM can't characterize the depth distribution of defects, XTEM has been performed to characterize defects throughout the top Ge layer depth range and down to Ge/Si interfaces. Sample U-5TC and LB-5TC represent the samples with low and high TDD in the Ge layers and were characterized by XTEM. Figure 6 shows the XTEM images of sample U-5TC and LB-5TC respectively. We can see that the TDs mainly exist in the Ge seeding layer (~100 nm thickness).

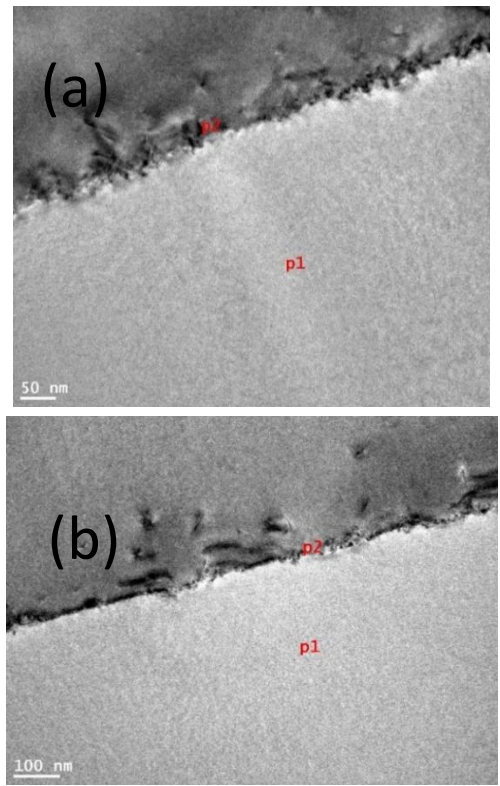

**Figure 6** Cross section TEM images in bright mode of Sample (a) U-5TC and; (b) LB-5TC. The TDD levels in Ge seeding layer of both samples are estimated to over $1\times 10^9$ cm$^{-2}$.

Despite the fact that U-5TC has a much lower TDD than LB-5TC in the top Ge layers measured by EPD which etched down to half of the top Ge layer thickness, both samples have an extremely high density of TDs in the Ge seeding layers, which are estimated to be over $1 \times 10^9$ cm$^{-2}$. These massive misfit dislocations were generated on the Ge/Si interfaces during growth to relax lattice-mismatch strain, as shown in HRXRD results in Section 2.3. Boron doping doesn't have much impact on the TDD in the Ge seeding layer according to the comparison between the two samples.

Considering the fact that U-5TC has the lowest TDD in the top Ge layers and LB-5TC has a much higher TDD than U-5TC, P-5TC and A-5TC, it is reasonable to assume that the TDDs of P-5TC and A-5TC in Ge seeding layer should also be between that of the U-5TC and LB-5TC, which is similar in this case. Thus, we conclude that all dopants had little impact on the TDD level in the Ge seeding layer, where the Si-Ge interdiffusion mainly happened.

### 3. Impacts of doping on Si-Ge interdiffusion
#### 3.1 Ge and dopants profiling

SIMS measurements were performed by Evans Analytical Group to obtain the Ge profiles in the samples. The samples were sputtered with 1 KeV Cs$^+$ primary ion beam obliquely incident on the samples at 60° off the sample surface normal. The sputter rate was calibrated using stylus profilometer measurements of total sputtered crater depths, and corrected on a point-by-point basis for the known sputter rate variation with SiGe composition. The measurement uncertainty in Ge fraction is ±1%.

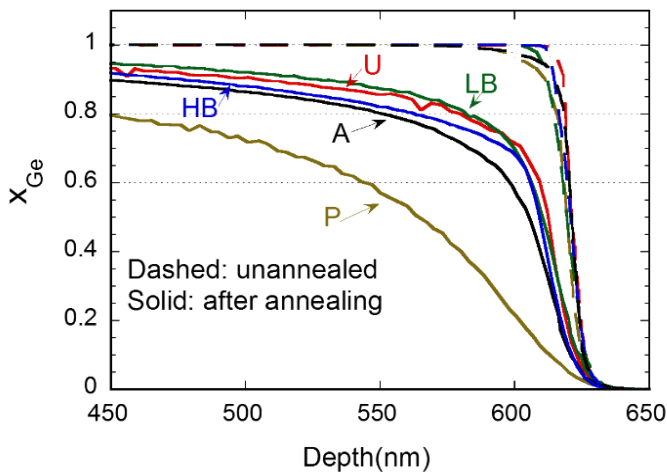

Figure 7 Ge profiles measured by SIMS. The dashed lines are the Ge profiles of samples without annealing. The solid lines are the Ge profiles of samples with annealing. The Ge profiles are shifted laterally for easy comparison.

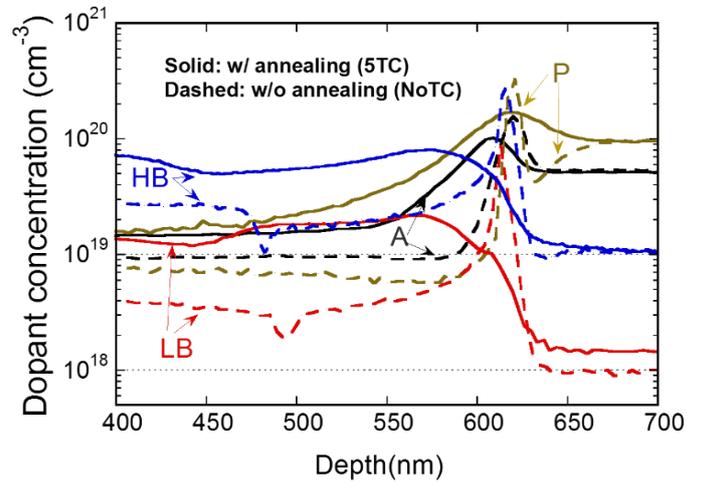

Figure 8 dopants (As/P/B) profiles of samples with and without annealing measured by SIMS.

Figure 7 and 8 show the Ge and dopants SIMS profiles respectively. As Sample LB-5TC has the same profiles as Sample LB-HT, only Sample LB-5TC is shown. Again, the 6 degree offcut difference and the merged annealing had no impact on the interdiffusion. The same is applied to Sample A-5TC and Sample A-HT. It is worth mentioning that the profiles from different samples have been shifted laterally. The reasons for doing that is because that Ge evaporation happened during the annealing reducing the Ge thicknesses of the annealed samples. Therefore, one cannot use the absolute depth of the SIMS profiles due to the thickness non-uniformity and depth errors discussed above. To compare the amount of interdiffusion, we use the slope of the Ge profiles as the evaluation criteria. Steeper Ge profiles mean less interdiffusion and vice versa.

In the Figure 7 and 8, we use dashed lines to stand for the samples without annealing, and solid lines for the samples with annealing. According to Figure 7, all samples have very similar sharp Ge profiles at the Ge/Si interfaces before annealing. The Ge profiles almost overlap with each other. All the dopants have the highest concentration at the interface of Ge/Si, which is due to the segregation induced by a high density of defects at the interfaces.

From Figure 7, we can see that sample U and sample LB have the least interdiffusion while sample P has the largest. Sample A has the second largest interdiffusion. While for sample HB, it has no significant difference over sample LB in $x_{Ge} < 0.7$ part, but it distinguishes itself from LB and U in $x_{Ge} > 0.7$ part. The interdiffusion profiles show a strong $x_{Ge}$ dependence, where much more diffusion happens in high Ge regions than in low Ge re-gions.

### 3.2 Effective interdiffusivity extraction

Boltzmann–Matano analysis was used to extract the time-averaged effective interdiffusivity ($\widetilde{D}_{Si-Ge}$) as a function of the Ge faction ($x_{Ge}$) from the concentration profiles. Theoretically, for Ge-Si interdiffusion couples, the condition for Boltzmann-Matano analysis is such that the interdiffusivity is only a function of the Ge molar fraction. It is not fully satisfied for extrinsic doping cases in P, As and highly boron doped samples as interdiffusivity depends on doping levels. On top of that, dopants diffuse and segregate during annealing while Si-Ge interdiffusion happens. Nevertheless, we can still use this method to estimate the interdiffusivities $\widetilde{D}_{Si-Ge}$. The meaning of the extracted time-averaged effective interdiffusivity $\widetilde{D}_{Si-Ge}$ is such that when this interdiffusivity is used in a finite difference time domain calculation, with the starting Ge profiles from the SIMS data of pre-annealed samples, the calculated post-annealing Ge profiles are consistent with the SIMS data of annealed samples.

The extracted $\widetilde{D}_{Si-Ge}$ was illustrated in Figure 9. Sample P has the highest interdiffusivity in the full $x_{Ge}$ range and the Sample A has the second largest interdiffusivity. The interdiffusivity of Sample P ($\widetilde{D}_P$) is 1.5 to 3 times higher than that of sample A ($\widetilde{D}_A$), and $\widetilde{D}_A$ is 1.5 to 2 times higher than that of Sample U ($\widetilde{D}_U$). Both As and P will enhance the interdiffusivity, which agrees with the previous study from our group [24, 25].

For P and As in Ge, as their diffusivities are much faster than Si-Ge interdiffusion, we can approximate that major dopant motion happens much faster than the major Si-Ge motion, which enables us to use the final dopant profiles as stable dopant distributions when most Si-Ge interdiffusion happens. This treatment is not valid for Ge with high boron doping, as boron diffusion in Ge (1.5 × $10^{-16}$ cm$^2$/s) [38] is comparable with Si-Ge interdiffusivity. In that case, the extraction of the time-averaged effective interdiffusivity ignores the boron doping effect.

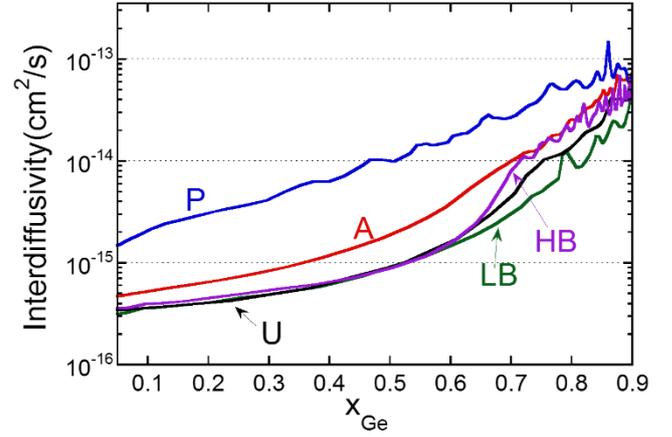

Figure 9 The time-averaged interdiffusivity as a function of Ge fraction using the Boltzmann-Matano method extracted from Sample U/P/A/HB/LB.

On the other hand, Sample LB, Sample HB, and Sample U do not exhibit much difference, especially in the $x_{Ge}$ < 0.6 part. In the $x_{Ge}$ > 0.6 range, their interdiffusivities do have some differences and it shows that $\widetilde{D}_{HB} > \widetilde{D}_U > \widetilde{D}_{LB}$. However, we should also keep in mind that we shouldn't over-interpret the difference. Due to the SIMS broadening effect and the data resolution limit, we estimate the error bar of our interdiffusivity extraction is +/-50%.

To conclude, n-type doping (As and P) can enhance the Si-Ge interdiffusivity significantly while boron's effect on that is small if any.

### 3.3 Mechanisms of interdiffusion enhancement

From the experimental results above, we can see that as long as high P/As doping exists, the Si–Ge interdiffusivity will increase significantly. To interpret this phenomenon, we need to discuss a few possible mechanisms:

1) Defect density. The possibility that this enhancement is due to a defect density difference is quite unlikely as the TDD values of Sample U and LB are close at Ge/Si interfaces. It is reasonable to assume that Sample P/A exhibit a similar result.

2) Strain. From our XRD results, Ge layers of all samples are under around 0.16% tensile strain from the coefficient of thermal expansion (CTE) mismatch. Therefore, all samples with different doping have similar strain levels, and the enhanced interdiffusion cannot be a result of strain difference.

3) Fermi-level effect. It is known that when a dopant concentration is close to or higher than the intrinsic carrier

concentration $n_i$, the Fermi-level effect can change the charged defect concentrations and thus the diffusivity [39]. The intrinsic carrier density of Si and Ge at T = 890 °C are 3.18 × 10$^{18}$ cm$^{-3}$ and 1.37 × 10$^{19}$ cm$^{-3}$ respectively, which are below the As/P doping concentration. This indicates that the Fermi-level effect existed during the annealing.

Cai *et al.* has investigated Si-Ge interdiffusion with a high P doping level by both experiments and modeling in the range of 0.75 < $x_{Ge}$ < 1 [25]. The doping dependence of Si-Ge interdiffusion was modeled successfully by a Fermi-enhancement factor. The ratio between extrinsic and intrinsic diffusion coefficient mediated by point defects can be expressed by the following formula:

$$\frac{\widetilde{D}(n)}{\widetilde{D}(n_i)} = \frac{1 + \sum_{r=1}^{2}\left(\frac{n}{n_i}\right)^r m_r \exp\left(\frac{rE_i - \sum_{n=1}^{r} E_{V^{n-}}}{kT}\right)}{1 + \sum_{r=1}^{2} m_r \exp\left(\frac{rE_i - \sum_{n=1}^{r} E_{V^{n-}}}{kT}\right)} \equiv FF$$

$(m_1 = 1, m_2 \geq 20).$  (1)

$E_i$ is the intrinsic Fermi level; $E_{V^-}$ and $E_{V^=}$ are energy levels of single negatively charged point defects ($V^-$) and doubly negatively charged point defects ($V^{2-}$) respectively. $m_1 = 1$ and $m_2 \geq 20$ show that interdiffusion is mainly dominated by $V^{2-}$ point defects. Due to limited literature resources of the energy levels of $V^-$ and $V^{2-}$ in SiGe, i.e. $E_{V^-}(x_{Ge})$, and $E_{V^{2-}}(x_{Ge})$, these terms was linearly interpolated between the value in Si and Ge, i.e.:

$$A_{r,SiGe}(x_{Ge}) = A_{r,Si}(1 - x_{Ge}) + A_{r,Ge}x_{Ge}$$
$(0 < x_{Ge} < 1),$  (2)

where $A_{r,SiGe}$ refers to the energy term $rE_i - \sum_{n=1}^{r} E_{V^{n-}}$ ($r \in \{1,2\}$) in SiGe. For Si, $A_{1,Si} = 0.1383$ eV and $A_{2,Si} = -0.1835$ eV [39]. For Ge, $A_{1,Ge} = -0.1134$ eV and $A_{2,Ge} = 0.0866$ eV. [40]

However, it will be inaccurate to use the same formula as our $x_{Ge}$ is from 0 to 1 instead of 0.75 to 1 in Cai et al.'s work. Interdiffusion can be mediated both by threading dislocations and by point defects in the lattice as seen in Eq. (3). The Fermi-enhancement factor shall only enhance $\widetilde{D}_{lattice}$.

$$\widetilde{D}_{total} = \widetilde{D}_{dislocation} + \widetilde{D}_{lattice}$$  (3)

The $\widetilde{D}_{dislocation}$ will dominate when $x_{Ge} < 0.5$, while $\widetilde{D}_{lattice}$ will dominate when $x_{Ge} > 0.7$ [23, 41, 42].

In Cai *et al.*'s study, $\widetilde{D}_{dislocation}$ was neglected as $x_{Ge} > 0.75$. In our case, we should modify $\widetilde{D}_{total}$ as the following equation:

$\widetilde{D}_{total} = \widetilde{D}_{dislocation} + \widetilde{D}_{lattice} * FF$, where

$$FF \equiv \frac{1 + \sum_{r=1}^{2}\left(\frac{n}{n_i}\right)^r m_r \exp\left(\frac{rE_i - \sum_{n=1}^{r} E_{V^{n-}}}{kT}\right)}{1 + \sum_{r=1}^{2} m_r \exp\left(\frac{rE_i - \sum_{n=1}^{r} E_{V^{n-}}}{kT}\right)}.$$  (4)

The $\widetilde{D}_{lattice}$ term can be calculated according to Ref. [20]. The $\widetilde{D}_{dislocation}$ term can be calculated as

$$\widetilde{D}_{dislocation} = \widetilde{D}(n_i) - \widetilde{D}_{lattice},$$  (5)

where $\widetilde{D}(n_i)$ is extracted from Sample U.

As for the calculation of electron density $n$, considering the charge neutrality equation $n = p + C_{P/As}$ and $n_i^2 = np$, the electron concentration n of the P-doped $Si_{1-x}Ge_x$ samples can be expressed as:

$$n(x_{Ge}, C_{P/As}) = \frac{C_{P/As} + \sqrt{C_{P/As}^2 + 4n_i^2(x_{Ge})}}{2}.$$  (6)

The last parameter is the calculation of $n_i(x_{Ge})$. Due to the limited data of $n_i(x_{Ge})$ at temperature over 600 °C, Cai *et al.* used linear interpolation between $n_{i,Ge}$ and $n_{i,Si}$ in his original code as Eq. (7). This approximation is good enough over the range 0.75 < $x_{Ge}$ < 1. However, our work covers 0 < $x_{Ge}$ < 1 range. The method used by Cai *et al.* overestimates $n_i(x_{Ge})$ when $x_{Ge}$ < 0.75. Since at $x_{Ge}$ < 0.85, $Si_{1-x}Ge_x$ alloys have been always considered as a "Si-like" material due to its band structure and electronic properties [43, 44]. Thus, we also tried exponential interpolation between $n_{i,Ge}$ and $n_{i,Si}$ as shown in (8).

$$n_i(x_{Ge}) = n_{i,Ge}x_{Ge} + n_{i,Si}(1 - x_{Ge})$$  (7)

$$n_i(x_{Ge}) = n_{i,Si}\exp(ln\frac{n_{i,Ge}}{n_{i,Si}} \times x_{Ge})$$  (8)

The simulation results with both approximation methods will be discussed in Section 3.4.

### 3.4 Simulation of Ge profiles after annealing

Since both sample LB-5TC and HB-5TC have similar Ge profiles with sample U-5TC, it means B has little impact on Si-Ge interdiffusion. We will focus our simulation work on sample A-5TC and P-5TC.

According to Eq. (4) and (6), the extrinsic interdiffusion coefficient increases with the concentration of dopant (P/As). However, as shown in Figure 8, during annealing, the concentration profile of dopant changed due to dopant diffusion and segregation. Ideally, it is best to simulate dopant diffusion, dopant segregation, and Si-Ge interdiffusion simultaneously. However, the diffusion and segregation of dopant involve many unknown coefficients such as the P/As diffusion segregation coefficients as a function of $x_{Ge}$, which are beyond the scope of this study. P and As diffusivity in Ge are over 20 times larger than $\widetilde{D}_{SiGe}$ [20, 45]. Therefore, we consider that most of the P and As motion happen in the early stage of the annealing process, and reach a distribution close to the final distribution. For Si-Ge interdiffusion, as it is much slower and fur-ther away from the equilibrium, we expect that the interdiffusion motion happens throughout the annealing. From that logic, in the annealing process, we expect that P and As reach a relative stable state much faster, and continue with small changes afterwards. Most of the interdiffusion motion happens after P and As reach a relative stable state, and we can approximate that the interdiffusion happens with a fixed P and As distribution same as the final profiles.

The simulation was done by Matlab™ to calculate $\widetilde{D}(n)$ in Eq. (4) and to simulate interdiffusion profiles using Fick's second law:

$$\frac{\partial C_{Ge}}{\partial t} = \frac{\partial}{\partial z}\left(\widetilde{D}(n)\frac{\partial C_{Ge}}{\partial z}\right). \quad (9)$$

To solve the diffusion equation numerically, we used finite difference time domain (FDTD) method. The experimental data of A-NA and P-NA were used as the initial profiles of the simulation. The boundaries are chosen to be far away from the interdiffusion region according to experimental data.

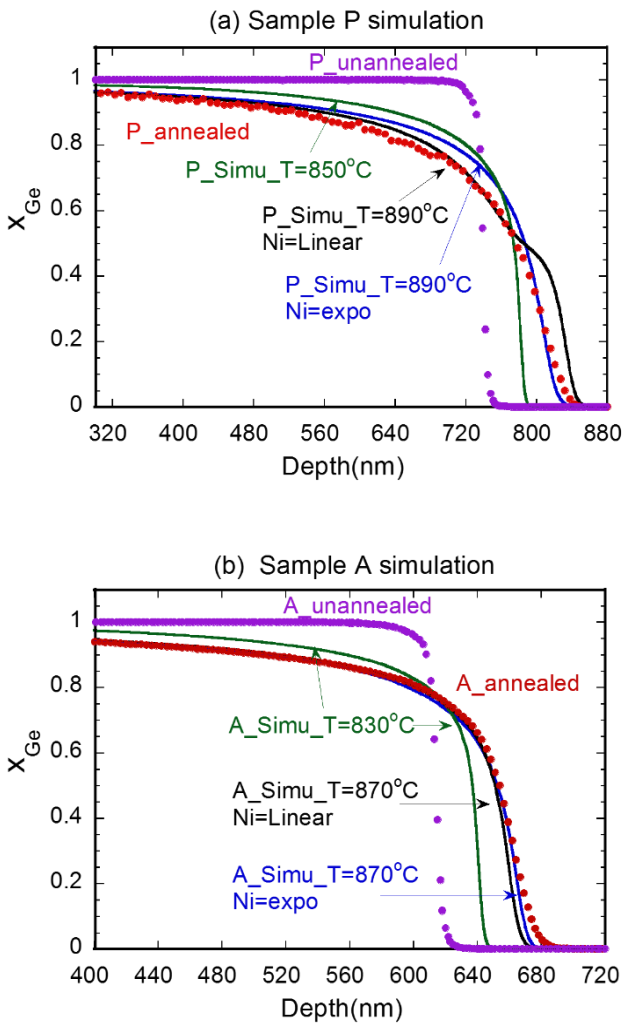

Figure 10 Simulation results (lines) with different parameters in comparison with SIMS data (symbols). (a) Sample P-5TC; (b) sample A-5TC. Temperature is set to be 890/850 °C for sample P and 870/850 °C for sample A. The $n_i(x_{Ge})$ is set to change exponentially or linearly with $x_{Ge}$.

As discussed early, the calibrated annealing temperate are 890 °C and 870 °C for Sample P-5TC and A-5TC respectively. We also calculated simulation at T = 850 °C for Sample P-5TC and at T = 830°C for Sample A-5TC to show the temperature sensitivity. The results are shown in Figure 11. Another critical parameter in simulation is $n_i(x_{Ge})$. Due to the limited data, two different models have been adopted. One is the exponential model based on Eq. (7), and the other is the linear model based on Eq. (8).

As illustrated in Figure 10, using T=890/870 °C and $n_i(x_{Ge})$ exponential model can give the best Ge profile fitting results for sample P-5TC/A-5TC respectively. According to Figure 10, although the linear interpolation of $n_i(x_{Ge})$ has a better fitting curve with P-5TC in $x_{Ge} > 0.6$ part compared with the exponential interpolation, it has a plateau in $0.4 < x_{Ge} < 0.6$. This plateau is not real, but a result due to the underestimation of interdiffusion enhancement in $x_{Ge} > 0.6$ region. Using the exponential interpolation of $n_i(x_{Ge})$ can solve this problem. Thus, we conclude that $n_i(x_{Ge})$ with an exponential dependence on $x_{Ge}$ is a better model than the $n_i(x_{Ge})$ with a linear dependence on $x_{Ge}$. As expected, the extracted temperatures (890/870 °C) worked better than the nominal reading temperature (850/830 °C) for Sample P-5TC/A-5TC.

At $x_{Ge} > 0.5$ part, P simulation results and P-5TC Ge SIMS profiles do not match as good. That inconsistence should be due to the over simplified treatment of P profile. According to the Kirkendall effect, as Si diffuses much faster in Ge than Ge in Si, more lattice sites move to the Ge side due to the unbalanced vacancy flux associated with the interdiffusion. These creation and annihilation of lattice sites result in the shift of P profile towards the Si side during the annealing procedure. In our simulations, we used the final P profile as the P distribution during interdiffusion. This is not the case when the Kirkendall effect is significant such that during interdiffusion, the true P profile is on the left side of the final P profile, which means that the true P concentration for a certain depth is higher than that what is read from the final P profile. Therefore, we underestimated the Fermi-enhancement effect at the Ge side when we treated P profile as unchanged. For sample A-5TC simulation results, we didn't observe a similar problem, which is likely due to the lower concentration of As and lower annealing temperature, making the Kirkendall effect less significant.

### 3.5 Observation of the Kirkendall effect

The Kirkendall effect is a very important effect for interdiffusion. According to the theoretical prediction, due to the much faster Si diffusion to the Ge side than Ge to Si, there will be net vacancy flux from Ge into Si. Therefore, the lattice sites originally on the Si side will reduce, and those on the Ge side will increase moving the Ge/Si interface towards the Si side. The creation and annihilation of lattice sites should result in the shift of dopants profiles, which can, in turn, influence the Si-Ge interdiffusivity. To our knowledge, there is currently no experimental observation of the Kirkendall effect in Si-Ge interdiffusion system due to the lack of inert marker elements in atomic scale. Nor was there any theoretical study on that.

To observe and study the Kirkendall effect, we annealed the P doped sample at T = 870 °C for 50 minutes and 200 minutes respectively in nitrogen gas. The reasons to use P doped Ge/Si were 1) faster interdiffusion to observe this effect with shorter annealing time and 2) P segregation peak at Ge/Si interface could be used as a marker layer to mark the movement of the Ge/Si interface.

To prevent the Ge sublimation during the annealing process, a 120 nm thick $SiO_2$ was deposited on Ge using plasma enhanced CVD at 300 °C before the annealing.

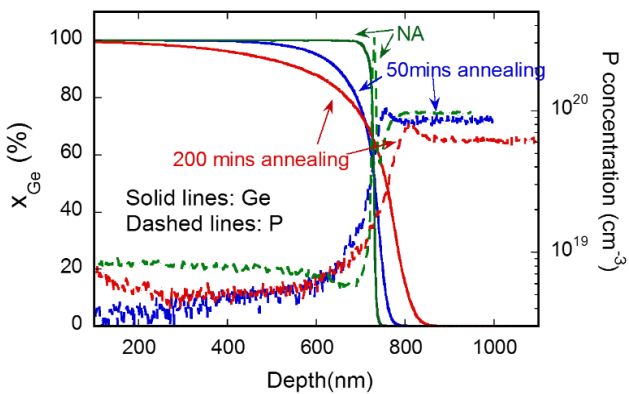

**Figure 11 Ge and P profiles of P doped samples with no annealing, with 50 minutes annealing and 200 minutes annealing respectively. The annealing was at T = 870 °C. The SIMS profiles have not been shifted as there is no Ge sublimation during annealing.**

Before annealing, P has a sharp segregation peak at the Ge/Si interface due to the high density of defects and dislocations. According to the SIMS data in Figure 11, compared to the unannealed sample P-NA, the lateral movement of P peak of P-NA after 200 min annealing is about 70 nm towards the Si side. This is larger than the possible depth error of ± 45 nm from the combination of the cross-wafer non-uniformity in epitaxial growth and the SIMS depth error. Therefore, we can confirm the observation of the Kirkendall effect, which lead to the lateral shifting of P during the annealing.

For doped Ge/Si structures, in fact, with only 3 elements such as P, Ge and Si, during thermal annealing, dopant diffusion, segregation, defect annealing, Si-Ge interdiffusion and exchange of lattice sites on both sides all happen simultaneously, making this a much more complicated physical picture than one expects.

## 4. Conclusions and technological implications

To summarize, this work studied Ge-on-Si growth and Si-Ge interdiffusion with different doping conditions both by experiments and theoretical modeling.

We found that different types of doping had no significant impact on the surface roughness and strain levels. All samples have a smooth surface (roughness < 1.5 nm). The Ge films are almost entirely relaxed. On the other hand, even a low B doping level introduces lots of extra TDs (TDD > $1 \times 10^8$ cm$^{-2}$) in the top Ge layers compared to the undoped samples. This greatly impairs the Ge films quality. The TDD value of undoped sample without annealing (U-NA) is $3.5 \pm 1.5 \times 10^6$ cm$^{-2}$. Although the annealing procedure can reduce the TDD to less than $10^5$ cm$^{-2}$ for undoped Ge films, it can be compromised by the interdiffusion between Si and Ge layers, which is highly undesired. Meanwhile, P and As doping can reduce the TDD values of Ge files to be less than $10^6$ cm$^{-2}$ without annealing. This offers a new method to fabricate high-quality Ge-on-Si films without defect annealing procedure, which is one of the major findings of this work. The two different annealing procedures (5TC and HT) and wafer offcut orientations (0 °C and 6 °C) have negligible impact on Ge film quality and interdiffusion. Photoluminescence studies of these samples and similar Ge films on (100) Si substrates are in process, and will be published separately.

Theoretically, it is the first time that Ge-Si interdiffusion with n-type doping that has been successfully simulated across the whole $x_{Ge}$ range. SIMS measurement results showed that Si-Ge interdiffusion is greatly enhanced in P and As doped samples, while B doping had little impact on that. We attributed this phenomenon to Fermi-level effect from P or As doping, which increases the negatively charged vacancy concentrations and thus the interdiffusivity. We used the extrinsic n-doped Si-Ge interdiffusion model with the Fermi-enhance factor to describe the impact of the Fermi-level effect. The validity of the model was proved by the comparisons between the simulations and the SIMS data from experiments with different anneal temperatures in $0 < x_{Ge} < 1$ range. We also reported the first observation of the Kirkendall effect in Ge-Si systems.

The results of this study are relevant to the design of optical and electronic devices with Ge-on-Si structures, including III-V lasers on Ge/Si, Ge-on-Si lasers, Ge modulators, and Ge photodetectors.

### Acknowledgements

This work was funded by Natural Science and Engineering Research Council of Canada (NSERC). This research was also supported by the National Research Foundation Singapore through the Singapore MIT Alliance for Research and Technology's Low Energy Electronic Systems (LEES) IRG, Competitive Research Program (NRF-CRP12-2013-04) and Innovation Grant from SMART Innovation Centre. Dr. Mario Beaudoin from the University of British Columbia are acknowledged for the training in HRXRD and EPD measurements and helpful discussions.